\documentstyle[11pt,newpasp,twoside]{article}
\input epsf

\def\eg{e.g.,\ }
\def\etal{et~al.\ }

\def\solar{\ifmmode_{\mathord\odot}\else$_{\mathord\odot}$\fi}

\def\edcomment#1{\iffalse\marginpar{\raggedright\sl#1\/}
\else\relax\fi}
\reversemarginpar

\begin{document}
\title{The Evolution of Tidal Debris}
\author{J. Christopher Mihos}
\affil{Case Western Reserve University, Department of Astronomy,
Cleveland, OH 44106}

\begin{abstract}
Galaxy interactions expel a significant amount of stars and gas
into the surrounding environment. I review the formation and evolution
of the tidal debris spawned during these collisions, and describe
how this evolution depends on the large scale environment in which the
galaxies live. In addition to acting as a long-lived tracer of the
interaction history of galaxies, the evolution of this material -- on
both large scales and small -- has important ramifications for
galactic recycling processes, the feeding of the intracluster light
and intracluster medium within galaxy clusters, and the delayed
formation of galactic disks and dwarf galaxies.
\end{abstract}

\section{The Physics of Tidal Tails}

The large-scale dynamical evolution of tidal debris is governed
largely by simple gravitational physics. As first elegantly shown in
simulations by Toomre \& Toomre (1972) and Wright (1972), the tidal
forces acting on spiral galaxies during a close encounter, coupled
with the galaxies' rotational motion, draw out long slender ``tidal
tails'' of gas and stars. An example of this process is shown in
Figure 1. As the galaxies pass by each other on the first passage,
tidal forces give disk material sufficient energy to escape the inner
potential well. The symmetric nature of tidal forces means that
streams are torn off both the near side and far side (with respect to
pericenter) of the disks; the near side material forms a tidal
``bridge'' between the disks (which typically does not physically
connect, depending on orbital geometry) while the far side material
forms the tidal tails. The formation of
tidal tails is a strong function of the orbital geometry -- tidal tails are
strongest in prograde encounters where the spin and orbital angular
momentum vectors are (even moderately) aligned, while retrograde
encounters yield weak tails at best. The length of the tidal tails is
further pronounced due to the orbital decay of the merging pair (\eg
Barnes 1988), which causes the galaxies to ``fall away'' from their
tails as they merge together.

\begin{figure}[h]
\plotfiddle{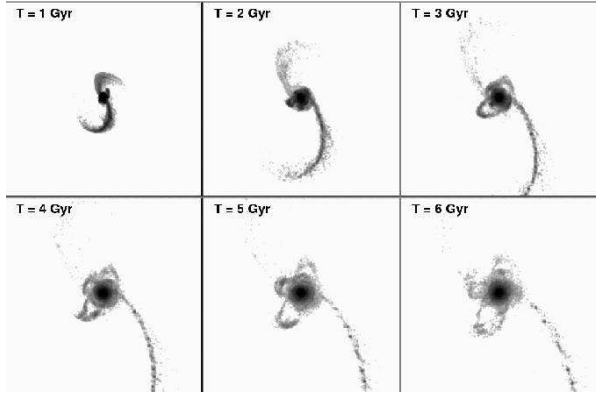}{2.0in}{0.}{40}{40}{-125}{0}
\caption{ Evolution of the tidal debris in an equal-mass merger of two
disk galaxies. Each frame is approximately 0.9 Mpc on a side. }
\end{figure}

Once launched, tidal tails are not in simple expansion. Figure 2 shows
the kinematic structure of the tidal tails shown in Figure 1, observed
1/2 Gyr after the merger is complete.  Most of the material remains
bound to the remnant on loosely bound elliptical orbits, with only the
relatively small fraction at the tip of the tails being unbound. The
radial velocity curve shows this orbital structure well: the loosely
bound outer portion of the tails are still expanding, while material
at the base of the tails has already reach apocenter and has started
falling back in towards the merger remnant. This velocity structure
results in a continual stretching of the tidal tails -- they are
long-lived and do not simply expand away, although their surface
brightness drops rapidly due to this dynamical evolution (Mihos
1995). One important caveat to this description is the depth of the
galaxies' potential well: a deep potential well provided by extended
dark matter halos will result in less unbound material and a more
rapid fall-back of the tidal debris to the parent galaxy (Dubinski
\etal 1996, 1999; Springel \& White 1999).

\begin{figure}[h]
\plotone{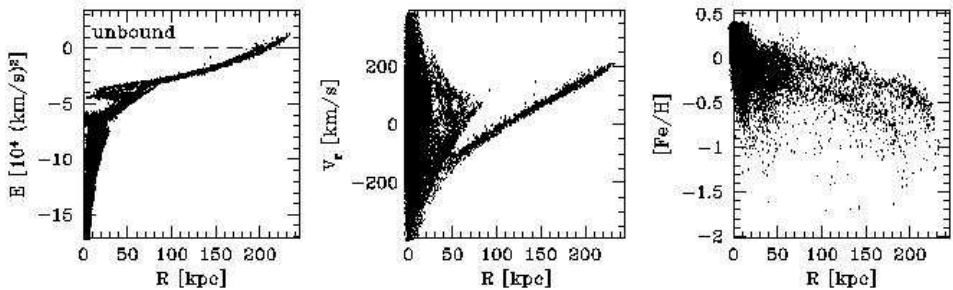}
\caption{The structure of the tidal debris from Figure 1, shown 0.5 Gyr 
after the merger is complete. Left: energy structure, middle: velocity
structure, right: metallicity distribution (see text).}
\end{figure}

The material forming the tidal tails comes from a wide range of
initial radii in the progenitor disks. During close passages, tidal
forces are effective at dredging up material from the inner disk and
expelling it into the tidal debris. In the simulation shown in Figure
1, scaled to Milky Way sized progenitors, the extended tidal tails are
formed from material originally outside the solar circle, while the
loops and shells which fall back in the first Gyr after the merger
include a significant amount of material from the solar circle and
inwards. This ``tidal dredge-up'' means that tidal debris will be
moderately metal rich, since it is not simply the outer parts of the
disks involved. To demonstrate this effect, we imprint a metallicity
distribution on the stellar disk model of d[Fe/H]/dR = $-0.05$ kpc$^{-1}
$, normalized to solar metallicity at the solar circle. Observed 1/2 Gyr
after the merger is complete, we see that a significant amount of the
debris in the outer (stellar) tidal tails has metallicities above 1/3
solar. A similar exercise for the gas skews the results towards lower
metallicities, for a number of reasons. Gas disks are typically more
extended than stellar disks; for a similar radial gradient there will
be more low metallicity material in the gas than the stars.
Additionally, the gas from the inner regions, which would have
provided higher metallicity gas in the tails, does not survive the
tidal expulsion process; instead shocks and gravitational torques
drive the gas inwards to the center of the remnant where it fuels the
merger-induced starburst instead (Mihos \& Hernquist 1996; Barnes \& Hernquist
1996).

\section{Galactic Recycling}

While on large scales the evolution of tidal debris is largely a
gravitational phenomenon, on smaller scales a variety of mechanisms
can drive structure formation within the tidal tails. Overdensities
can form in the tidal tails either through gravitational collapse of
small scale instabilities in the progenitor disks (Barnes \& Hernquist
1992) or by cooling and fragmentation of structure in the tidal
expelled gas (Elmegreen \etal 1993).  This has led to the suggestion
that dwarf galaxies may form within the tidal debris of merging
galaxies. Observations have detected a number of discrete, often
star-forming, sources in the tidal debris of interacting galaxies
(\eg Duc, these proceedings);
whether or not these are truly bound objects destined to become dwarf
galaxies remains to be seen.

We can use simulations of interacting galaxies to make predictions for
the properties of any tidally spawned dwarfs. Coming from material
stripped from their progenitor disks, they should have moderate
metallicities and travel on loosely bound, highly eccentric orbits
(Hibbard \& Mihos 1995). They are unlikely to have significant
amounts of dark matter, since the kinematically hot dark matter will not
collapse into the shallow potential wells (Barnes \& Hernquist 1992)
formed from small-scale instabilities in the tails.
Finally, these tidal dwarfs may well show different generations of
stellar populations, as they arise in a mixed medium of old stellar
disk material and young stars formed from the gaseous tidal debris.

The dynamical stretching of the tidal debris means that it should be
hard for these condensations to grow continuously. On small scales,
bound structures can form, but continual accretion onto these
structures will be limited by shear in the surrounding
material. In this context, it is important to make a cautionary note
about claims that large, tidally spawned HI complexes are
often found preferentially at the end of optical tidal tails.
Dynamically it is unclear why this would be -- HI tails often extend
much further out than the optical tails do, and there is not clear
reason why the ``end of the optical tails" should be a dynamically
important spot. It is more likely that many of these objects are the
result of projection effects. Tidal tails are curved, and a sightline
which passes along the tangent point to a curving tail will not only
give the appearance of marking the end of the tail, but also will
project along a large column of HI, artificially giving the impression
that a massive HI complex lives at the end of a tidal tails (see \eg
Hibbard, these proceedings, but also Bournaud \etal 2003 for an
alternative view).

The other context in which tidal debris is important in galactic
recycling is the return of gas from the infalling tidal debris. As
shown in \S 1, material in the tidal tails remains bound, and will
continue to fall back to the remnant over many Gyr. The return is
ordered (Hibbard \& Mihos 1995); the first material to return is the
most bound, lowest angular momentum material, which will fall back to
small radius. As the remnant evolves, high angular momentum, loosely
bound material will fall back to increasingly larger radius.

This long-lived ``rain'' of tidal debris on the merger remnant
manifests itself in a number of ways. Diffuse loops and shells form as
the stars fall back through and wrap around the remnant, while the
infalling gas can dissipate energy and settle into a warped, rotating
disk (Mihos \& Hernquist 1996; Naab \& Burkert 2001; Barnes 2002),
such as those found in the nearby elliptical galaxies NGC 4753
(Steiman-Cameron \etal 1992) and Centaurus A (Nicholson \etal 1992).
The most loosely bound tidal material forms less-well organized
structures outside of a few effective radii as it falls back, and may
be the source of the extended HI gas found in shells and broken rings
around many elliptical galaxies (\eg Schminovich \& van Gorkom 1997). More speculatively, if the returning gas can efficiently form stars,
this process provides a mechanism for rebuilding stellar disks. For
example, the gaseous disk inside the merger remnant NGC 7252 is
rapidly forming stars (Hibbard \etal 1994), and may ultimately result
in a kiloparsec-scale stellar disk embedded in the $r^{1\over 4}$
spheroid formed in the merger. If significant amount of tidal material
exists to reform a stellar disk, it may even be possible for the
remnant to eventually evolve towards a bulge-dominated S0 or Sa galaxy
(\eg Schweizer 1998).

\section{Tidal Debris in Clusters}

Many dynamical avenues are available to drive tidal evolution in
cluster galaxies. The most obvious is the cluster potential itself,
particularly for galaxies whose orbit takes them close to the cluster
center (\eg Henriksen \& Byrd 1996). More recently, the importance of
repeated, fast collisions in stripping cluster galaxies has been
emphasized by Moore \etal (1996, 1998). However, because of the large
velocity dispersion within galaxy clusters, conventional wisdom held
that strong interactions and mergers between cluster galaxies were
rare (Ostriker 1980).

More recently, a greater understanding of the nature of hierarchical
clustering is changing this view. While slow encounters are rare for
an individual galaxy falling into a well-established environment (\eg
Ghigna \etal 1998), many galaxies are accreted onto clusters from
within the small group environment. Clusters show ample
evidence for substructure in X-rays, galaxy populations, and velocity
structure (see, \eg reviews by Buote 2002; Girardi \& Biviano 2002).
Interactions within infalling
groups can be strong -- witness, for example, the classic interacting
pair ``the Mice'' (NGC 4676), found in the outskirts of the Coma
cluster. Clearly strong interactions can and do occur during the evolution
of clusters, either early as the cluster forms, or late as groups
are accreted from the field.

The effects of the cluster potential on the evolution of tidal debris
during a slow encounter can be dramatic. To illustrate this, Fig 3
shows the evolution of the same merger shown in Fig 1,
except this time occurring in a Coma-like cluster potential. The orbit
of the galaxy pair in the cluster carries it within 0.5 Mpc of the
cluster core, with an orbital period of $\sim$ 3.5 Gyr. As the galaxies
merge, the very loosely bound material forming the tidal tails is now
subjected to the large scale tidal field of the cluster, and is very
efficiently stripped out of the galactic potential altogether.

\begin{figure}[h]
\plotfiddle{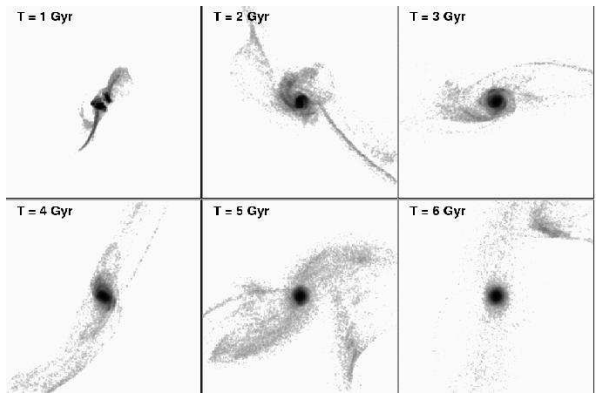}{2.0in}{0.}{40}{40}{-125}{0}
\caption{Evolution of an equal-mass merger, identical to that in
    Fig. 1, but occurring as the system orbits through a
    Coma-like cluster potential (see text).  Note the rapid stripping of
    the tidal tails early in the simulation; the tidal debris seen
    here is more extended and diffuse than in the field merger, and
    late infall is shut off due to tidal stripping by the cluster
    potential.
}
\end{figure}

An extremely important facet of this kind of encounter is the enhanced
efficiency of the tidal stripping. This is shown in Figure 4, which
shows the fraction of material stripped to large radius ($r>35$ kpc, or
approximately 5 $R_e$ in the simulation) in the field and cluster
versions of the simulations, as well as in a single disk galaxy on the
same cluster orbit. The combination of the local and cluster tides
causes significant stripping -- encounters of galaxies in  small infalling
groups effectively ``prime the pump" for the
cluster tides to do their work. Indeed, the {\it individual} disk galaxy is
hardly tidally stripped at all, suggesting that estimates of tidal
stripping based on the tidal radius of individual galaxies falling
into a cluster potential may significantly underestimate the effect.

\begin{figure}[h]
\plotone{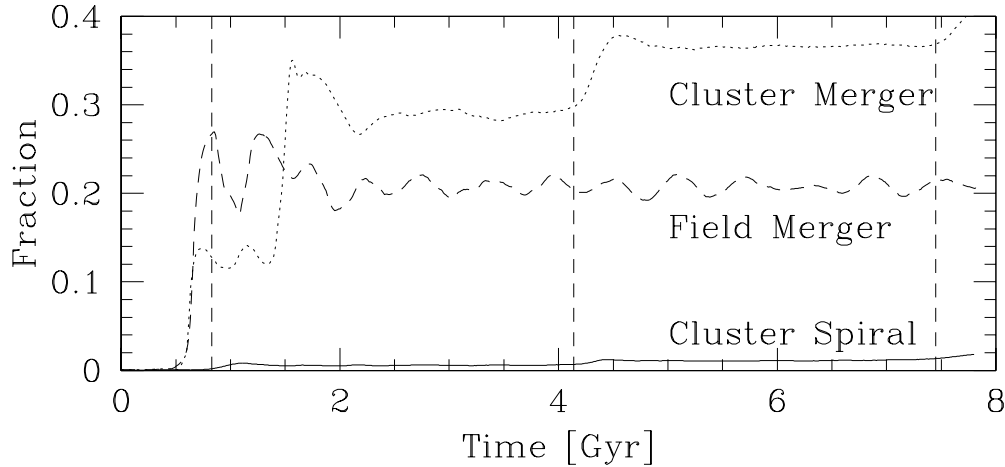}
\caption{Material stripped to large radius ($r>35$ kpc) for the isolated
merger, the cluster merger, and a isolated spiral orbiting in the cluster
potential. Cluster peri passages are shown as dashed lines. Note that most
of the tidal material in the isolated merger remains bound to the remnant,
while it is unbound in the cluster merger.}
\end{figure}

The combined effects of galactic and cluster tides not only raise the
efficiency of tidal stripping, they also result in particularly {\it
deep} stripping. That is, the stronger galactic tides can strip
material out from deep in the galaxies' potential well, which is then
vulnerable to the gentler but long-lived cluster tides that liberate
it entirely. As a result, the stripped material will be relatively
high in metallicity, coming from the inner parts of the disk, and has
a mean metallicity of [Fe/H]=$-0.25$, with a significant spread. This
has important consequences for studies of the intracluster light
(ICL), particularly in terms of searches for individual intracluster
stars which are sensitive to the metallicity of the population (\eg
Durrell etal 2002).

In terms of galactic recycling, the cluster has the effect of
essentially shutting down various recycling paths. The ability for
tidal tails to grow large tidal dwarfs may be extremely limited, as
the cluster tides rapidly disperse the tidal material. The hot
intracluster medium may also act to heat the tidal gas, making it
difficult to form stars. If any dwarfs or, on smaller scales, star
clusters do form in the tidal debris, they will be rapidly stripped
from their hosts, perhaps contributing to the populations of cluster
dwarfs or intracluster globular clusters.

The cluster will also shut down reaccretion from the tidal tails
spawned during a merger. The combination of cluster tides and ram
pressure stripping from a hot intracluster medium will ``sweep clean"
the tidal debris and any low density gas that might remain in the
remnants. For example, the diffuse HI disk in the merger remnant
Centaurus A (Nicholson \etal 1992) is unlikely to survive any passage
through the hot ICM of a dense cluster. Models for forming S0 galaxies
from mergers of galaxies followed by reformation or survival of a
gaseous disk (\eg Bekki 1998, or see the discussion in Schweizer 1998)
seem difficult to envision in the dense cluster environment. However,
the S0 classification is a very diverse one, and the mechanism which
gives rise to disky cluster S0's may well be quite different than the
merger mechanisms hypothesized to give rise to bulge-dominated S0's in
the field environment.

\section{The Formation of Intracluster Light}

As galaxies orbit in the cluster environment, they are subject to
tidal stripping from a variety of sources -- interactions with
individual galaxies, with groups of galaxies, or with the global
cluster potential itself (see, \eg the discussion in Gnedin 2003).
Over time, this stripped starlight builds up the diffuse intracluster
light found in clusters of galaxies. The properties of this ICL -- its
luminosity, morphological structure, metallicity, and kinematics --
and their correlation with cluster properties can help unravel the
dynamical history of cluster collapse, accretion, and evolution.
To date, theoretical work has largely focused on tidal stripping from
individual galaxies orbiting in an evolved cluster potential (\eg
Merritt 1983; Richstone \& Malamuth 1983; Moore \etal 1996;
Calc\'aneo-Rold\'an \etal 2000) and ignored two important effects:
preprocessing in groups, and heating by substructure (Gnedin 2003).
Full cosmologically-motivated simulations are needed to study the
phenomenon in detail (\eg Dubinski \etal 2001; Napolitano \etal 2003;
Mihos \etal 2004).

An example of these models is shown in Figure 5 (from Mihos \etal
2004). In this simulation, we have excised a cluster from a flat
$\Lambda$CDM cosmological simulation and traced it back to
$z=2$. At that point we identify dark matter halos more massive than $10^{11}$
M$_{\solar}$ which destined to end up in the $z=0$ cluster and replace
them with composite (collisionless) disk/halo galaxy models.  The
simulation is then run forward to the present day to examine the
formation of tidal debris and the ICL. In essence, this simulation
follows the contribution to the ICL from luminous galaxies, rather
than from the stripping of low mass dwarfs. In this simulation, we see
significant kinematic and spatial substructure at early times; at late
times much of this substructure has been well mixed into a diffuse
intracluster light. However, at low surface brightnesses, significant
substructure remains even at $z=0$.

\begin{figure}[h]
\plotfiddle{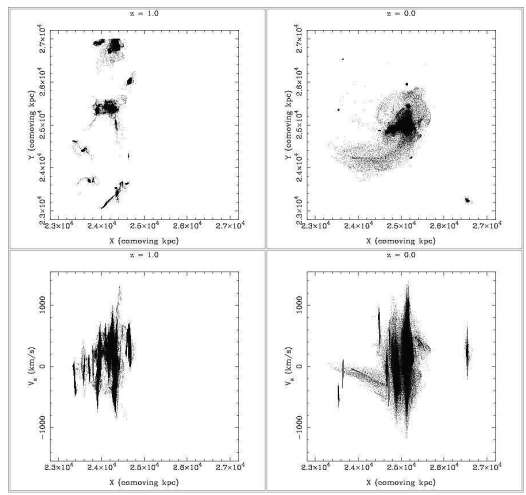}{3.0in}{0.}{70}{70}{-130}{0}
\caption{Morphological (top) and kinematic (bottom) structure of the
intracluster light in a simulated galaxy cluster. Left panels show the
cluster at $z=1$, while the right panels show $z=0$. From Mihos \etal (2004).}
\end{figure}

Detecting this ICL has proved difficult, as at its {\it brightest},
the ICL is only $\sim$1\% of the brightness of the night sky. Efforts
to detect this ICL include deep surface photometry to look for the
diffuse ICL (\eg Uson \etal 1991; Bernstein \etal 1995; Gonzalez \etal
2000; Feldmeier \etal 2002), as well as imaging of individual stars
and planetary nebulae in nearby clusters (Ferguson \etal 1998;
Feldmeier \etal 1998; Arnaboldi \etal 2002). Recently, these surveys
have begun to reveal interesting substructure in the ICL, often in the
form of diffuse arcs or streaks of material from tidally stripped
galaxies (Trentham \& Mobasher 1998; Gregg \& West 1998;
Calc\'aneo-Rold\'an \etal 2000).

To quantify the prevalence and properties of ICL as a function of
cluster properties, we have begun a deep imaging survey of clusters
using the KPNO 2m (Feldmeier \etal 2002, 2004). We target a variety of
clusters, from cD-dominated Bautz-Morgan Type I clusters to Type III
clusters which are typified by a more irregular distribution of
galaxies. Examples from this survey are shown in Figure X. The massive
cD cluster Abell 1413 is marked by regular distribution of diffuse
light, well-fit by a $r^{1\over 4}$ distribution over a large range of
radius, with only a moderate excess at large radius and little
substructure.  In contrast, Abell 1914 shows a variety of features: a
fan-like plume projecting from the eastern clump of galaxies, another
diffuse plume extending from the galaxy group to the north of the
cluster, and a narrow stream extending to the northwest from the
cluster center. We see similar behavior in other Abell clusters we
have surveyed.

\begin{figure}[h]
\plottwo{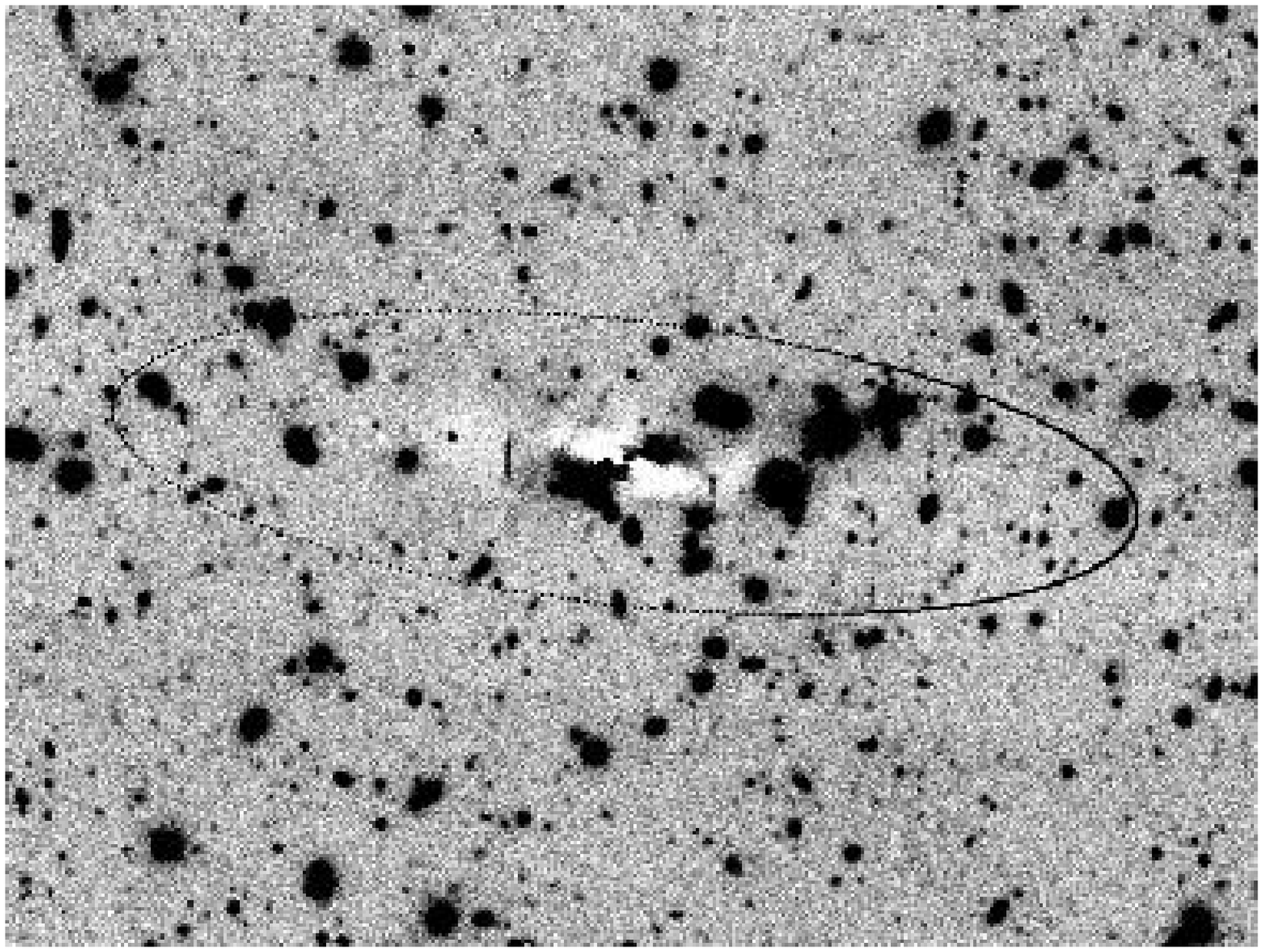}{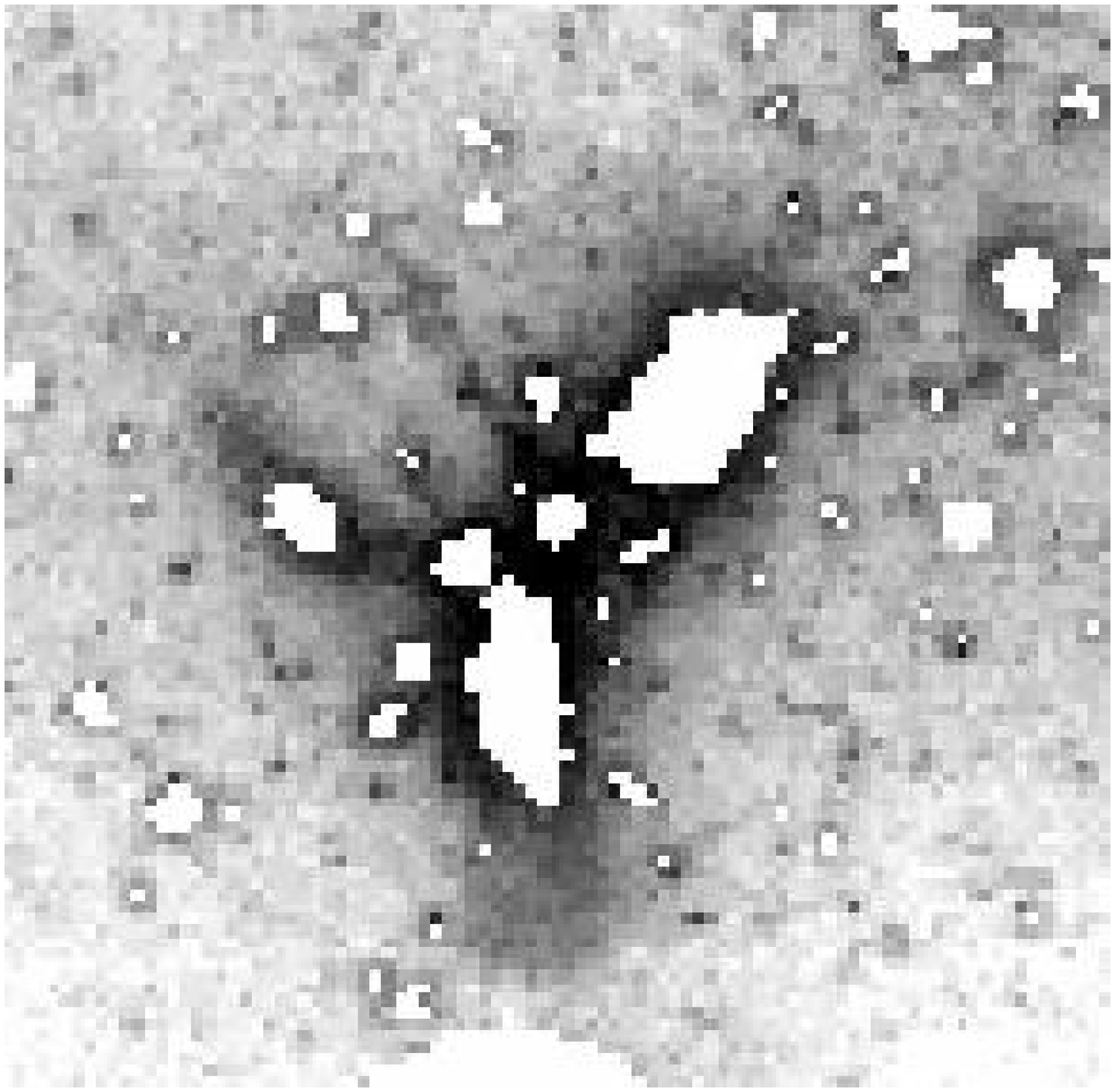}
\caption{Left: the cD cluster Abell 1413, after subtraction of a smooth
$r^{1\over 4}$ law (the extent of which is shown by the ellipse). Very little
substructure is seen. Right: the Bautz-Morgan Type III cluster Abell 1914,
showing a rich variety of substructure. North is to the left; east is down.
(From Feldmeier etal 2002, 2004)}
\end{figure}

Although the sample size is small, these results are consistent with
the expectations that substructure in the ICL is correlated with the
dynamical state of the cluster as a whole. As clusters are assembled,
the ICL is built up though the significant tidal stripping that occurs
during interactions within the accreting groups, and between galaxies
and substructure within the cluster. Does the total amount of ICL
also correlate with Bautz-Morgan cluster type? Examining ICL measurements
from a variety of sources, Ciardullo \etal (this conference)
find only a weak dependence -- the ICL fraction rises as expected from
Type III to Type II clusters, but Type I (cD-dominated) clusters show
fractionally less ICL than do the Type II's. However, the drop in the
Type I's is likely due to the difficulty in distinguishing the ICL
from the diffuse envelope of the cD galaxy itself; indeed, such 
distinction may not even be well motivated, since the cD envelope
itself likely is formed from tidally stripped material. Including the
luminosity of the cD envelope in the ICL budget would raise the fractional
amount of ICL in Type I clusters and bring the trend in line with
expectations from the dynamical models for generating ICL in clusters.

\acknowledgments{My numerous collaborators have all made many
contributions to this work. In particular, I thank Cameron McBride for
his work generating and visualizing the cluster ICL simulations. This
work has been supported in part by the NSF through a CAREER award
AST-9876143 and by a Research Corporation Cottrell Scholarship.}

\end{document}